# FDDM: Frequency-Decomposed Diffusion Model for Rectum Cancer Dose Prediction in Radiotherapy

Xin Liao, Zhenghao Feng, Jianghong Xiao, Xingchen Peng, and Yan Wang, *Member*, *IEEE*

*Abstract*—Accurate dose distribution prediction is crucial in the radiotherapy planning. Although previous methods based on convolutional neural network have shown promising performance, they have the problem of over-smoothing, leading to prediction without important high-frequency details. Recently, diffusion model has achieved great success in computer vision, which excels in generating images with more high-frequency details, yet suffers from time-consuming and extensive computational resource consumption. To alleviate these problems, we propose Frequency-Decomposed Diffusion Model (FDDM) that refines the high-frequency subbands of the dose map. To be specific, we design a Coarse Dose Prediction Module (CDPM) to first predict a coarse dose map and then utilize discrete wavelet transform to decompose the coarse dose map into a low-frequency subband and three high-frequency subbands. There is a notable difference between the coarse predicted results and ground truth in high-frequency subbands. Therefore, we design a diffusion-based module called High-Frequency Refinement Module (HFRM) that performs diffusion operation in the high-frequency components of the dose map instead of the original dose map. Extensive experiments on an in-house dataset verify the effectiveness of our approach.

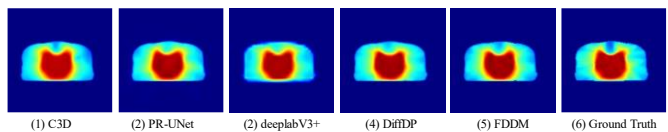

Fig. 1. Examples from a rectum cancer patient. (1) to (5) prediction from C3D [12], PRUNet, deeplabV3+, DiffDP [13], and FDDM.

## I. INTRODUCTION

RADIOTHERAPY, as a crucial therapy in cancer treatment, plays a pivotal role in improving patients' survival rates and their quality of life. Radiotherapy Planning (RP) is an important process, aiming to ensure adequate therapeutic dosage to the planning target volume (PTV) while minimizing potential harm to the organs at risk (OARs). To obtain a clinically acceptable RP, dosimetrists need to manually adjust the optimization objectives in a trial-and-error manner. This process is time-consuming and labor-intensive [1]. Besides, the quality of the dose map heavily relies on dosimetrists' expertise and experience [2]. Therefore, it is important to develop methods that automatically predict dose distribution maps for patients.

Recently, due to the rapid development of deep learning (DL), especially the convolutional neural network (CNN), many CNN-based dose prediction methods have been proposed [3-8]. Kearney et al. [3] proposed a 3D UNet-based model called DoseNet to predict the dose distribution map. Nguyen et al. [4] employed a modified 2D UNet [9] for dose prediction in prostate cancer patients. Wang et al. [5] added a progressive refinement module in the UNet (PRUNet) to produce more accurate dose distributions.

Song et al. [6] utilized the deepLabV3+ [10] to extract contextual information from different scales, improving the accuracy of dose distribution prediction for patients with rectum cancer. Wang et al. [7] disassembled the dose distribution into many beam paths to transform the task into several easy tasks. Furthermore, following the idea of Generative Adversarial Network (GAN), Zhan et al. [8] introduced a dual attention module to predict a more precise dose map (Mc-GAN).

Although these CNN-based methods have achieved promising performance, they often suffer from the over-smoothing problem [11], leading to predictions without essential high-frequency details. We display some predicted dose maps from five different DL models in Fig. 1, from Fig. 1, we can see predictions from (1) to (3) lack high-frequency information. These high-frequency details reflect information

This work is supported by National Natural Science Foundation of China (NSFC 62371325, 62071314), Sichuan Science and Technology Program 2023YFG0263, 2023YFG0025, 2023NSFSC0497, and Opening Foundation of Agile and Intelligent Computing Key Laboratory of Sichuan Province.

Corresponding author: Yan Wang.
Xin Liao, Zhenghao Feng, and Yan Wang are with the School of Computer Science, Sichuan University, China. (e-mail: liaoxin0528@163.com; fzh_scu@163.com; wangyanscu@hotmail.com).
Xingchen Peng is with the Department of Biotherapy, Cancer Center, West China Hospital, Sichuan University, China. (e-mail: pxx2014@scu.edu.cn).
Jianghong Xiao is with the Department of Radiation Oncology, Cancer Center, West China Hospital, Sichuan University, China. (email: xiaojh86@foxmail.com).

on the direction of radiation passing through the human body, which is crucial for protecting OARs and killing as many cancer cells as possible. In recent years, denoising diffusion probabilistic model (DDPM) [14] has achieved good performance in various computer vision tasks, Feng et al. [13] introduced a structure encoder into the DDPM (DiffDP) to better guide the DDPM to generate the dose map. In Fig. 1, we find that DDPM is good at generating high-frequency details [15], but it requires long running times and substantial computational resources to achieve iterative denoising [16].

In this paper, to tackle the problem, we propose the Frequency-decomposed Diffusion Model (FDDM) that leverages the strength of diffusion models for high-frequency details to obtain an accurate dose map and improve the efficiency of diffusion models. Specifically, we first gain a coarse dose map by our proposed Coarse Dose Prediction Module (CDPM). Then, we decompose the coarse dose map into four subbands with three high-frequency subbands and a low-frequency subband by discrete wavelet transform (DWT), which significantly reduces the spatial dimension without compromising information [17]. Accordingly, we propose a High-Frequency Refinement Module (HFRM) that performs diffusion operation in high-frequency subbands rather than the original dose map, resulting in a faster speed compared to the previous diffusion-based method, DDPM. Furthermore, to better guide HFRM to generate the high-frequency components of the dose map, we condition HFRM at both the image level and feature level. Finally, we use inverse wavelet transform (IWT) for dose map reconstruction. Our contributions to this work can be summarized as follows: 1) we propose a dose prediction framework called Frequency-decomposed Diffusion Model (FDDM) for rectum cancer patients, which harnesses the generative ability of diffusion models on high-frequency details and the strengths of the discrete wavelet transform (DWT) for more precise and efficient dose prediction. 2) we design a Coarse Dose Prediction Module (CDPM) to generate the coarse map and a High-Frequency Refinement Module (HFRM) that performs diffusion operation in the wavelet domain to refine the high-frequency components of the dose map. 3) we conduct substantial experiments on an in-house dataset comprising 130 rectum cancer patients, demonstrating the outstanding performance of our proposed method.

## II. METHODS

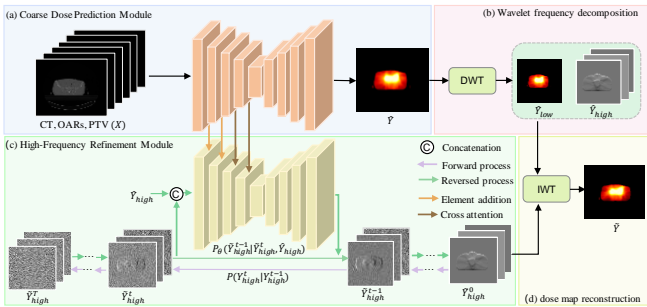

Fig. 2. Overview of the proposed FDDM network. 'DWT' in (b) and 'IWT' in (d) represent discrete wavelet transform and inverse wavelet transform, respectively.

An overview of the proposed FDDM model is illustrated in Fig. 2, we first produce a coarse dose map by Coarse Dose Prediction Module (CDPM) and do discrete wavelet transform (DWT) to resolve the coarse prediction into three high-frequency subbands and a low-frequency subband. Then we introduce a High-Frequency Refinement Module (HFRM) to generate refined high-frequency components. Finally, we utilize inverse wavelet transformation (IWT) to reconstruct the predicted dose map.

### A. Coarse Dose Prediction Module

To make the predicted dose map encompass as rich information as the ground truth, we propose a Coarse Dose Prediction Module (CDPM), shown in Fig. 2(a). CDPM takes X as input and outputs a coarse dose map $\hat{Y}$, where X includes the original CT images and the segmentation masks of the PTV and OARs. Encouraged by the great success of UNet [10], we utilize a six-level UNet to construct CDPM. Specifically, the encoder involves five modules, where each module contains a Residual Block (ResBlock) and a Downsampling block, except for the final one. The ResBlock includes two convolutional blocks (ConvBlock), each containing a 3 × 3 convolutional (Conv) layer, a GroupNorm (GN) layer, and a Swish activation function. To address the problem of gradient vanishing during training, we introduce residual connections. Furthermore, the Downsampling block consists of a 3 × 3 Convolutional layer with a stride of 2, allowing for effective downsampling of the feature maps. The decoder includes five deconvolution blocks to upsample, where each block comprises an Upsampling block and two Resblocks with the last one excluding the Upsampling block, and each Upsampling block utilizes the Nearest neighbor up-sampling and a Conv layer with a kernel size of 1 to implement the upsampling operation. Additionally, a bottleneck is incorporated between the encoder and decoder, which consists of two ResBlocks and a self-attention module. The skip connections are utilized to enable multi-level feature reuse and aggregation between the encoder and decoder. During training, we use L1 loss to minimize the content difference between the coarse prediction $\hat{Y}$ and ground truth $Y$:

$$L_{CDPM} = |\hat{Y} - Y|. \quad (1)$$

### B. Wavelet Frequency Decomposition

The output of the CDPM is a coarse prediction, which is blurred with a few high-frequency details. To solve this problem, we refine the high-frequency subbands of the coarse dose map. Specifically, we use 2D discrete wavelet transformation (DWT) with Haar wavelets [18] to decompose the coarse dose map $\hat{Y} \in R^{H \times W}$ into four high- and low-frequency subbands, illustrated in Fig. 2(b). This process can be formulated as:

$$DWT(\hat{Y}) = \{\hat{Y}_{LH}, \hat{Y}_{HL}, \hat{Y}_{HH}, \hat{Y}_{LL}\}, \quad (2)$$

where $\hat{Y}_{LH}, \hat{Y}_{HL}, \hat{Y}_{HH}, \hat{Y}_{LL} \in R^{\frac{H}{2} \times \frac{W}{2}}$, $\hat{Y}_{high} = \{\hat{Y}_{LH}, \hat{Y}_{HL}, \hat{Y}_{HH}\}$ represents the high-frequency subbands of the dose map, which contains the edge and detail information, and $\hat{Y}_{low} = \{\hat{Y}_{LL}\}$ is the low-frequency subband of the dose map, encompassing the overall structural information.

## C. High-Frequency Refinement Module

The coarse prediction is over-smoothed without many high-frequency details, leading to significant differences compared to the ground truth on high-frequency subbands. To generate a dose map with more details, we propose High-Frequency Refinement Modules (HFRM), shown in Fig. 2(c). Our proposed HFRM is based on the Denoising Diffusion Probabilistic Model (DDPM) [14], which consists of two processes: the forward and the reverse process. Different from traditional DDPM, we conduct diffusion operations in the wavelet domain instead of the image space, which allows for accelerated inference speed. Consequently, the forward process in HFRM first pred-defines a variance schedule $\{\beta_1, \beta_2, \dots, \beta_T\}$ to progressively transform high-frequency components of the dose map $Y_{high}^0$ to a noisy data $Y_{high}^T \sim \mathcal{N}(0, I)$ through T steps:

$$q(Y_{high}^t | Y_{high}^{t-1}) = \mathcal{N}(Y_{high}^t; \sqrt{1-\beta_t} Y_{high}^{t-1}, \beta_t I). \quad (3)$$

Besides, based on the property of the Markov Chain, we can directly get $Y_{high}^t$ at any intermediate timestep t from $x_0$ through the formula:

$$q(Y_{high}^t | Y_{high}^0) = \mathcal{N}(Y_{high}^t; \sqrt{\bar{\alpha}_t} Y_{high}^0, (1-\bar{\alpha}_t)I)$$
$$= \sqrt{\bar{\alpha}_t} Y_{high}^0 + (1-\bar{\alpha}_t)\epsilon_t, \quad (4)$$

where $\bar{\alpha}_t \coloneqq \prod_{i=1}^t (1-\beta_i), \epsilon_t \sim \mathcal{N}(0, I)$.

In the reverse phase, the HFRM generates images by denoising a randomly sampled standard Gaussian noise $\tilde{Y}_{high}^T \sim \mathcal{N}(0, I)$ into clear high-frequency subbands $\tilde{Y}_{high}^0$, the process is also a Gaussian transition. In a classical diffusion model, this process is usually conditioned in the image level, namely simply concatenating the struct image as an image prior. However, the reverse process beginning with a sampled randomly Gaussian noise may suffer from a content inconsistency between the predicted results and ground truth owing to the diversity of the sampling process. Therefore, we employ conditions in both the image and feature levels:

$$p_\theta(\tilde{Y}_{high}^{t-1} | \tilde{Y}_{high}^t) = \mathcal{N}(\tilde{Y}_{high}^{t-1}; \mu_\theta(\hat{Y}_{high}, X, \tilde{Y}_{high}^t, t), \sigma_t^2 I), \quad (5)$$

where $\sigma_t^2$ is usually a fixed variable related to $\beta_t$, and $\tilde{Y}_{high}^t$ are high-frequency components of coarse prediction, and $\mu_\theta(\hat{Y}_{high}, X, \tilde{Y}_{high}^t, t)$ is usually predicted by a denoising UNet $\epsilon_\theta(\hat{Y}_{high}, X, \tilde{Y}_{high}^t, t)$:

$$\mu_\theta(\hat{Y}_{high}, X, \tilde{Y}_{high}^t, t) = \frac{1}{\sqrt{\alpha_t}} \left( \tilde{Y}_{high}^t - \frac{1-\alpha_t}{\sqrt{1-\bar{\alpha}_t}} \epsilon_\theta(\hat{Y}_{high}, X, \tilde{Y}_{high}^t, t) \right). \quad (6)$$

The UNet architecture in our proposed HFRM shares the same structure as the CDPM. At the image level, we concatenate high-frequency components $\hat{Y}_{high}$ and intermediate result $\tilde{Y}_{high}^t$ as input. For the feature level, feature maps of CDPM are element-wise added at the first two encoder modules, while in the following three encoder modules, we apply a cross-attention operation to gain similarity-based structure guidance at deeper levels.

The main goal is to optimize the denoising model $\epsilon_\theta$ to achieve a close approximation between the predicted noise $\epsilon_\theta(\hat{Y}_{high}, X, \tilde{Y}_{high}^t, t)$ in the reverse process and added noise $\epsilon_t$ in the forward process, the training objective can be formulated as:

$$L_{HFRM} = E_{Y_{high}^0, t, \epsilon_t \sim \mathcal{N}(0,I)}[\|\epsilon_t - \epsilon_\theta(\hat{Y}_{high}, X, \tilde{Y}_{high}^t, t)\|]. \quad (7)$$

## D. Dose Map Reconstruction

After gaining refined high-frequency subbands, we utilize IWT on the refined high-frequency subbands and the low-frequency subband of the coarse dose map to reconstruct the dose map. This process is shown in Fig. 2(d) and can be formulated as:

$$\tilde{y} = IWT(\tilde{Y}_{high}^0, \hat{Y}_{low}). \quad (8)$$

The whole model is constructed in an end-to-end mode and the total loss $L_{total}$ can be formulated as the sum of (1) and (7):

$$L_{total} = L_{CDPM} + L_{HFRM}. \quad (9)$$

## III. EXPERIMENTS

### A. Dataset and Evaluation Metrics

To evaluate the performance of our proposed method, we conduct extensive experiments on a dataset curated from an in-house rectum cancer dataset at West China Hospital. It contains 130 patients who underwent volumetric modulated arc therapy (VMAT) treatment. Every patient contains CT images, OAR segmentations, PTV segmentation, and clinically planned dose distribution. In addition, OAR segmentations include the small intestine (ST), FemoralHeadL(FHL), FemoralHeadR (FHR), and bladder (BLD). We randomly select 98, 10, and 22 patients for training, validation, and testing, respectively. Before training, we split the 3D volumes with a resolution of 3mm×3mm×3mm into multiple 2D slices with a size of 160 × 160 along the axial direction. Finally, we obtained 16346 slices for training, 1529 slices for validation, and 3491 slices for testing.

We select multiple metrics for the evaluation of our proposed model. We employ $D_2, D_{50}$, mean dose ($D_{mean}$), and conformity index (CI) [19] for evaluation. Here Dm denotes the lowest dose absorbed that encompasses a volume percentage of m% of the PTV. CI is defined as:

$$CI = \frac{(V_{PTV} \cap V_{100\%ISO})^2}{V_{PTV} \times V_{100\%ISO}}, \quad (10)$$

where $V_{PTV}$ represents the volume of the PTV, and $V_{100\%ISO}$ indicates the isodose volume, which is a binary mask that is assigned a value of 1 when the dose value of a voxel exceeds the threshold value, and 0 otherwise. We calculate the difference (Δ) of these metrics between the predicted results and ground truth. Furthermore, we employ the dose volume histogram (DVH) [20] as an intuitive metric for comparison.

TABLE I
QUANTITATIVE COMPARISON RESULTS WITH STATE-OF-THE-ART METHODS ON PTV

| Methods | $|\Delta CI|$ | $|\Delta D_2|$(Gy) | $|\Delta D_{50}|$(Gy) | $|\Delta D_{mean}|$(Gy) |
|---|---|---|---|---|
| UNet [4] | 0.131(0.136) | 1.043(0.383) | 2.221(1.468) | 2.077(1.352) |
| U-ResNet-D [21] | 0.0868(0.127) | 1.043(0.383) | 2.556(1.684) | 2.416(1.504) |
| deepLabV3+ [6] | 0.0650(0.142) | 0.547(0.462) | 0.975(1.593) | 0.968(1.450) |
| C3D [12] | <u>0.0635(0.144)</u> | 0.561(0.476) | 0.801(1.388) | 0.856(1.185) |
| PRUNet [5] | 0.0651(0.136) | 0.754(0.422) | 0.914(1.638) | 0.959(1.465) |
| Mc-GAN [8] | 0.0827(0.150) | 1.004(0.458) | 1.273(1.215) | 1.268(1.049) |
| DiffDP [13] | 0.0683(0.131) | <u>0.486(0.350)</u> | <u>0.776(1.353)</u> | <u>0.780(1.180)</u> |
| Proposed | **0.0626(0.137)** | **0.309(0.192)** | **0.754(1.484)** | **0.762(1.308)** |

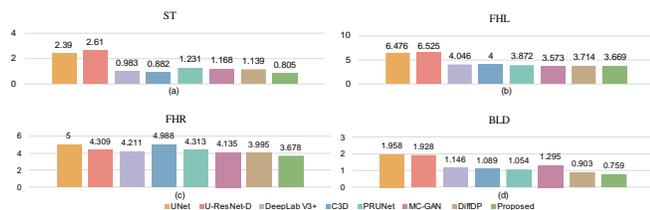

Fig. 3. Quantitative results of SOTA methods on OARs (i.e., ST, FHL, FHR, and BLD) in terms of $|\Delta D_2|$.

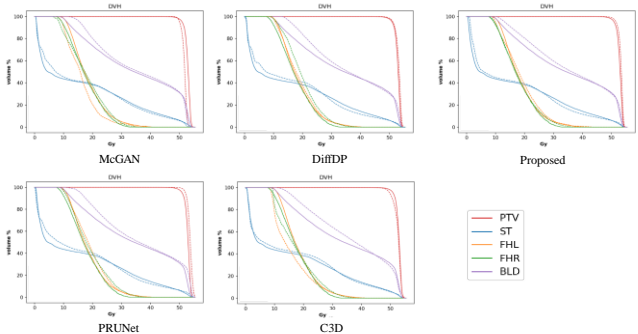

Fig. 4. DVH curves for comparison with SOTA methods. The solid line denotes the ground truth while the dotted line denotes the predictions.

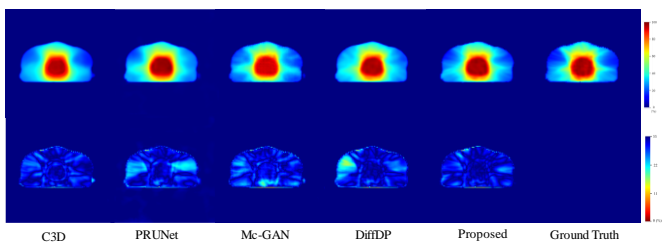

Fig. 5. Visualization of the proposed method and SOTA methods. The first row visualizes the predicted dose maps and the the second row represents the error heat maps.

When DVH curves of the ground truth and predictions are closer, we can infer better prediction results.

*B. Implementation Details*

Our proposed method is implemented with the PyTorch framework on an NVIDIA RTX 3090 GPU with 24 GB memory. We use the Adam as optimizer to train the model for 1300 epochs with a learning rate of 1e−4 and a batch size of 16. Additionally, the time step number of diffusion model T is set to 1000 during training.

*C. Comparison with State-Of-The-Art Methods*

We compare the performance of the proposed method to the state-of-the-art (SOTA) methods, including UNet [4], U-ResNet-D [21], deepLabV3+ [6], C3D [12], PRUNet [5], Mc-GAN [8], and DiffDP [13] for comparison. The quantitative comparison results are listed in Table. 1, illustrating the superiority of our method over existing SOTA models in all metrics. Specifically, compared with DiffDP with the second-best accuracy in $|\Delta D_{50}|$ (0.776 Gy) and $|\Delta D_{mean}|$ (0.780 Gy), the results generated by the proposed are 0.022 and 0.018 lower, respectively. As for $|\Delta D_2|$, method outperforms the second-best approach by a margin of 0.309 Gy. Furthermore, we also display some quantitative results on $|\Delta D_2|$ of OARs in Fig. 3, which demonstrates our proposed method achieves the best performance on the OARs. Besides, we also present the DVH curves in Fig. 4. Compared with other methods, the disparity between the DVH curves of our method and the ground truth is the smallest, especially for ST and BLD, demonstrating the superior performance of the proposed. Furthermore, we visualize the prediction results in Fig.5. As we can see, the dose map generated by our proposed method contains more high-frequency details. Besides, the error map of our proposed method appears to be the darkest, indicating the least disparity when compared to the ground truth.

TABLE II
ABLATION STUDY OF OUR METHOD ON PTV IN TERMS OF $\Delta CI$, $\Delta D_2$, $\Delta D_{50}$, AND $\Delta D_{mean}$

| | $|\Delta CI|$ | $|\Delta D_2|$(Gy) | $|\Delta D_{50}|$(Gy) | $|\Delta D_{mean}|$(Gy) |
|---|---|---|---|---|
| (A) | 0.0805(0.142) | 0.886(0.408) | 0.913(1.383) | 0.911(1.224) |
| (B) | 0.0637(0.125) | 0.804(0.385) | 0.855(1.505) | 0.909(1.351) |
| (C) | 0.0659(0.141) | 0.490(0.266) | 0.810(1.531) | 0.831(1.366) |
| (D) | **0.0626(0.137)** | **0.309(0.192)** | **0.754(1.484)** | **0.762(1.308)** |

*D. Ablation Study*

To verify the effectiveness of pivotal components in our proposed method, we conduct a series of ablation experiments. The experiment arrangement can be generally concluded as (A) only CDPM, (B) only HFRM (use HFRM to directly predict the dose map), (C) we use a CDPM to generate a coarse dose map and use another CDPM to refine the high-frequency subbands, and (D) CDPM+HFRM (the proposed method). we compare the experimental results between (B) and (D), compared to (B), (D) exhibits a substantial improvement in all four metrics. Particularly, in terms of CI, in which (D) outperforms (B) by a margin of 0.495, demonstrating its effectiveness.
The Experimental results are given in Table 2. We can compare the results of (A) and (D), after adding HFRM, $\Delta CI$, $\Delta D_2$, $\Delta D_{50}$, and $\Delta D_{mean}$ improve a lot with 0.0179, 0.577 Gy, 0.159 Gy, and 0.149 Gy, respectively. Besides, we can see that (D) outperforms (C) in all metrics, illustrating the generality ability of high-frequency details of the HFRM.

## IV. CONCLUSION

In this paper, we present Frequency-Decomposed Diffusion Model (FDDM) for robust and efficient rectum cancer dose prediction in radiotherapy. Technically, we design a Coarse Dose Prediction Module (CDPM) to gain a coarse dose map. Besides, we proposed a High-Frequency Refinement Module (HFRM) that conducts the diffusion process in the wavelet space rather than the image space to generate a more accurate dose map. Moreover, we condition our proposed HFRM in both the image and feature level for better guidance. The experimental results demonstrate that our method significantly outperforms other methods.